\documentclass[a4paper,12pt]{article}
\usepackage[english]{babel}
\usepackage{times}
\usepackage{latexsym} 
\usepackage{amsmath}
\usepackage{amssymb}
\usepackage{graphicx}
\usepackage{wrapfig}
\usepackage{indentfirst}
\makeatletter
\def\@seccntformat#1{\csname the#1\endcsname.\quad} \makeatother

\newcommand{\be}{\begin{equation}}
\newcommand{\ee}{\end{equation}}
\newcommand{\ba}{\begin{eqnarray}}
\newcommand{\ea}{\end{eqnarray}} 

\let\f\frac

\begin{document}
\title{\large\textbf{The Field Theory of Gravitation and The Rest Mass of Particles}}
\date{}

\author{\normalsize S.\,S.\, Gershtein, A.\,A.\, Logunov and M.\,A.\, Mestvirishvili}

\maketitle

\begin{abstract}

{It is shown in this work that all free physical fields should have a \textbf{nonzero rest mass} according to the
field theory of gravitation.}.

\end{abstract}

The Relativistic Theory of Gravitation (RTG), as a field theory, considers the gravitational field as a physical
field with spins $ 2 $ and $ 0 $, propagating in the Minkowski space. The source of this field is a universal
conserving quantity --- the energy-momentum tensor of all fields of substance, including the gravitational field
also. Just such an approach to gravitation leads to the effective Riemannian space of a field origin. Let us note
that the effective Riemannian space may have only trivial topology. The motion of a test body occurs in the
Minkowski space under action of the gravitational force. This is equivalent to the motion of a test body along a
geodesic line in the effective Riemannian space. In the framework of such an approach  there is the following
complete system of  RTG equations derived from the least action principle [1, 2]: \be
R^{\mu\nu}-\f{\,1\,}{2}g^{\mu\nu}R+\f{m^2}{2} \Bigl[g^{\mu\nu}+\Bigl(g^{\mu\alpha}g^{\nu\beta}
-\f{\,1\,}{2}g^{\mu\nu}g^{\alpha\beta}\Bigr)\gamma_{\alpha\beta}\Bigr] =8\pi T^{\mu\nu}\,, \label{eq1} \ee \be
D_\nu\tilde{g}^{\mu\nu}=0\,. \label{eq2} \ee As the gravitational field acts in the Minkowski space with  metric
tensor $ \gamma_{\mu\nu} $, it should not allow the motion of a  test body to proceed outside  the  Minkowski
space null-cone. This is ensured by the  causality requirements: \be g_{\mu\nu}U^\mu U^\nu =0\,, \label{eq3} \ee
\be \gamma_{\mu\nu}U^\mu U^\nu \geq 0\,. \label{eq4} \ee Here $U^\mu$ is the \textbf{isotropic} velocity 4-vector
in the effective Riemannian space corresponding to the physical fields having \textbf{zero rest mass}.

A \textbf{time-like} velocity 4-vector in the Riemannian space satisfying the following equation
\[
g_{\mu\nu}U^\mu U^\nu =1,\quad U^\nu = dx^\nu/ds\,,
\]
where $ds$ is the interval of effective Riemannian space, corresponds to the physical fields having
\textbf{nonzero rest mass}.

The particle 4-momentum is defined by a well-known equality
\[
p^\nu = mcU^\nu.
\]
According to causality conditions (\ref{eq3}) and (\ref{eq4}) any time-like velocity vector in the effective
Riemannian space \be g_{\mu\nu}U^\mu U^\nu =1 \label{eq5} \ee should stay time-like in the Minkowski space also,
i.e. \be \gamma_{\mu\nu}U^\mu U^\nu >0\,. \label{eq6} \ee Due to the fact that conditions  (\ref{eq3}) and
(\ref{eq4}) should be fulfilled, in particular, for the weak gravitational fields, we obtain, according to the
perturbation theory, \be g_{\mu\nu}=\gamma_{\mu\nu}-\phi_{\mu\nu}+\f{\,1\,}{2}\gamma_{\mu\nu}\phi, \quad \phi
=\gamma_{\mu\nu}\phi^{\mu\nu}. \label{eq7} \ee For a weak gravitational field, for example, for a weak
gravitational wave, condition  (\ref{eq3}) takes the following form: \be \gamma_{\mu\nu}U^\mu U^\nu
=\phi_{\mu\nu}\underset{0}{U}^\mu \underset{0}{U}^\nu\,. \label{eq8} \ee Here the following equation is taken into
account: $\gamma_{\mu\nu}\underset{0}{U}^\mu \underset{0}{U}^\nu =0$. The r.h.s. of Eq.~(\ref{eq8}) is not
positive definite, and so it is possible that condition  (\ref{eq4}) is violated. Just due to this reason it is
necessary to exclude any chance for the following equation to be valid \be g_{\mu\nu}U^\mu U^\nu =0 \label{eq9}
\ee for any field, because this   contradicts to the causality requirements. These requirements should be valid
for all physical fields due to the universal nature of the gravitational field.

A discussion concerning the causality principle had place in articles~[3--6]. But it had not fully removed
objections, stated in papers~[3--4]. In order to exclude any possibility for violation of the causality principle
it is necessary to formulate a general physical conclusion: \textbf{all free physical fields, including
electromagnetic one, have a nonzero rest mass}. This general physical conclusion from the RTG is in good
correspondence with the basic Minkowski axiom [7]:
 \textit{``A substance being at any worldpoint can be always considered as staying at rest under reasonable
 definition of space and time}.\\ \hspace*{5mm}{\bf\textit{The axiom tells us in other words that at any worldpoint
 the following expression {\mathversion{bold} $c^2dt^2-dx^2-dy^2-dz^2$} is always positive or in other words that any
 velocity
{\mathversion{bold}$v$} is always less than {\mathversion{bold}$c$}\,. According to this, {\mathversion{bold}$c$}
is the upper limit for supersubstantial velocities and this is a more profound meaning of quantity
{\mathversion{bold}$c$}}}\,''.

On the base of our general physical conclusion the causality conditions (\ref{eq3}) and (\ref{eq4}) are reduced to
the following conditions \be g_{\mu\nu}U^\mu U^\nu =1\,, \label{eq10} \ee \be \gamma_{\mu\nu}U^\mu U^\nu > 0\,.
\label{eq11} \ee Just these conditions were given by us in article [5], whereas Eq.~(\ref{eq8}) was ignored.

In case of a weak gravitational field we find from the above \be \gamma_{\mu\nu}U^\mu U^\nu =1+\underset{0}{U}^\mu
\underset{0}{U}^\nu \Bigl(\phi_{\mu\nu}-\f{\,1\,}{2}\gamma_{\mu\nu}\phi\Bigl)>0\,. \label{eq12} \ee Here
$\underset{0}{U}^\nu =dx^\nu/d\sigma$, $d\sigma$ is the Minkowski space interval \be
\gamma_{\mu\nu}\underset{0}{U}^\mu \underset{0}{U}^\nu =1\,. \label{eq13} \ee Therefore, according to
Eq.~(\ref{eq12})  vector $U^\nu$ being time-like in the effective Riemannian space stays time-like also in the
Minkowski space. This means that the null-cone of the effective Riemannian space is disposed inside the cone of
the Minkowski space. So, constant $c$, entering the expression for Minkowski space interval \be d\sigma^2
=c^2dt^2-dx^2-dy^2-dz^2 \label{eq14} \ee is a universal constant which unites space and time into a unified
space-time continuum. It always stays the unattainable upper limit for the velocity of motion of any kind of
matter. The fact that this conclusion follows from the RTG can be explained in the following way: due to
universality of gravity its requirements should be fulfilled by all free physical fields.

In conclusion authors express their gratitude to V.\,A.\,~Petrov, A.\,N.\,~Tavkhelidze, N.\,E.\,~Tyurin and
Yu.\,V.\,~Chugreev for valuable discussions.

\end{document}